\documentclass[aip,pop]{revtex4-1}
\usepackage{dcolumn,bm,amsmath,amssymb} 
\usepackage[pdftex]{graphicx}

\usepackage{color}
\usepackage{natbib}
\usepackage{amssymb}
\usepackage{amsmath}
\usepackage{xcolor}

\usepackage{subcaption}

\DeclareMathOperator{\arcsinh}{arcsinh}

\begin{document}


\title{Gravitational waves generated by laser accelerated relativistic ions}
\author{Evgeny G. Gelfer}\email{egelfer@gmail.com}
\affiliation{National Research Nuclear University ``MEPhI'' (Moscow Engineering Physics Institute), Kashirskoe sh. 31, 115409 Moscow, Russia}
\affiliation{Institute of Physics of the ASCR, ELI-Beamlines, 18221 Prague, Czech Republic}
\author{Hedvika Kadlecov\'{a}}\email{Hedvika.Kadlecova@eli-beams.eu}
\affiliation{Institute of Physics of the ASCR, ELI-Beamlines,  18221 Prague, Czech Republic}
\author{Ond\v{r}ej Klimo}\affiliation{Institute of Physics of the ASCR, ELI-Beamlines, 18221 Prague, Czech Republic}
\affiliation{FNSPE, Czech Technical University in Prague, 11519 Prague, Czech Republic}
\author{Stefan Weber}\affiliation{Institute of Physics of the ASCR, ELI-Beamlines, 18221 Prague, Czech Republic}
\author{Georg Korn}\affiliation{Institute of Physics of the ASCR, ELI-Beamlines, 18221 Prague, Czech Republic}

\begin{abstract}
The generation of gravitational waves by laser accelerated relativistic ions is investigated. The piston and light sail models of laser plasma acceleration are considered and analytical expressions for space-time metric perturbation are derived. For both models the dependence of gravitational waves amplitude on the laser and plasma parameters as well as gravitational waves spectrum and angular distribution are examined. \end{abstract}


\maketitle

\section{Introduction}
The remarkable progress in high power laser technology raises the question, how the expected tens of petawatt \cite{ELI,VULCAN} or even exawatt \cite{XCELS} 
laser facilities can be used for fundamental research. One of the widely discussed suggestions is strong field quantum electrodynamics (QED), which becomes observable
 \cite{Sauter1931,Schwinger1951,Bulanov2010,Bell2008,Fedotov2010,Gelfer2015,Narozhny2015}.

In this paper we discuss another problem, the use of high power lasers for investigation of gravity. Namely, we study ions, accelerated by ultra strong laser pulse, 
as a source of gravitational waves (GW)\cite{Maggiore}, a prediction of Einstein's general theory of relativity.

From the 1960s, there were attempts to detect GW coming from space first with Weber resonant detectors \cite{Weber1960} in the frequency range $<100$ Hz, 
later with interferometers such as LIGO \cite{abramovici1992,abbott2009} or VIRGO \cite{acernese2006} in the frequency range $10$ Hz to $20$ kHz. 
Later, the so called GW Hertz experiments were considered, which consist of the generation and detection of GW under laboratory conditions using for GW generation nuclear explosion \cite{chapline1974} or particle accelerator \cite{chen1991}, see more examples in \cite{li2002,rudenko2004}. 

Recently it was suggested  to use ions accelerated by laser field as terrestrial generator of GW\cite{Ribeyre2012}. 
Such waves would emit GW in the frequency range of GHz to THz.
In  forthcoming laser facilities \cite{ELI,VULCAN,XCELS} the speed 
of accelerated ions would be close to the speed of light, and in this paper a relativistic approach is used to investigate GW. 

GW from the astrophysical event, merging of two black holes, were registered recently by LIGO collaboration \cite{abbot}. However the generation of GW in the laboratory would provide outstanding opportunities for investigation of this phenomenon. In the present paper we will study the possibility of such generation with forthcoming laser facilities. 

Gravitational waves have also been linked to high power lasers in the recent paper \cite{Han2014}, where GW from electron-positron pairs, created by strong electromagnetic field were considered. 

Since GW are very weak, the linear approximation of Einstein equations is used
\begin{equation}
g_{ij}=\eta_{ij}+h_{ij},\quad \Box h_{ij}=-\frac{16\pi G}{c^4}T_{ij}, 
\end{equation}
where $\Box =\frac{1}{c^2}\frac{\partial^2}{\partial{t^2}}-\Delta$ is the d'Alembert operator, $T$ is the energy-momentum tensor, $g$ is the metric tensor, 
$\eta$ is the metric of the Minkowski spacetime, and $|h_{ij}| \ll 1$ is the perturbation of the flat spacetime, caused by gravitational waves, $G$ is the gravitational 
constant and $c$ is the speed of light. We use Gaussian units throughout. The gravitational waves are assumed to be plane waves because of the large distance from the source as compared to wavelength. 

We consider two models of ion acceleration. First of them is the piston model \cite{Naumova2009,Schlegel2009,Macchi2013,Qiao2009}, 
which describes laser acceleration of a thick plasma target. 
Electrons are pushed forward by the radiation pressure force and induce a charge separation in the plasma, generating
 a strong electrostatic field, which accelerates ions. As a result a shock wave like structure is formed in the plasma \cite{Naumova2009}. 
 The velocity of the shock wave front in the piston reference frame coincides with the piston velocity in the laboratory frame and is equal \cite{Naumova2009} to 
\begin{equation}\label{vf}
v_f=\beta_fc=\frac{B}{1+B}c,\quad B=\sqrt{\frac{I}{m_{i}n_{i}c^3}},
\end{equation}
where $I$ is the laser pulse intensity, $m_{i}$ the ion mass, and $n_{i}$ the ion density.

If the plasma target is thin, then radiation pressure can completely separate charges and further accelerate ions, because they are not anymore screened by the 
background plasma \cite{Macchi2013}. This regime is described by the Light Sail (LS) model 
\cite{Macchi2013,Esirkepov2004,Robinson2008,Macchi2009,Qiao2009}. The equation of motion of the target reads
\begin{equation}\label{eqls}
\frac{du^{j}}{dt}=\frac{2I}{m_{i}n_{i}lc^2}\frac{\sqrt{1+u^2}-u^{j}}{\sqrt{1+u^2}+u^{j}},
\end{equation}
where $u^{j}$ is the component of the 4-velocity along the direction of the laser pulse propagation, and  $l$ is the thickness of the target. 
The optimal for acceleration thickness of the target can be estimated as \cite{Macchi2009}:
\begin{equation}\label{l}
l\sim \frac{a_0}{\pi}\frac{n_c}{n_e}\lambda,\quad a_0=\sqrt{\frac{I}{n_cm_ec^3}},
\end{equation}
where $\lambda$ is the laser wavelength and $n_e$ is the initial electron density,  $n_c=m_e\omega^2/4\pi e^2$ is the critical density, 
$m_e$ and $e$ are electron mass and charge, and $\omega$ is the laser frequency $\omega=2\pi c/\lambda$.

\section{ GW amplitude}
\subsection{Piston model}
In the laser piston case accelerated ions move along the laser pulse direction at the velocity \cite{Naumova2009,Schlegel2009}
$v_i=2\beta_f c/(1+\beta_f^2)$.
Suppose, that a laser pulse propagates along the $x$ direction. 
Then the non-vanishing spatial component of the energy-momentum tensor \cite{LL2} reads 
\begin{equation}\label{emt}
T^{xx}(t,\mathbf{r})=\rho_0 c^2(u^{x})^2\Theta(t,\mathbf{r}),
\end{equation}
where $\rho_0$ is the mass density of accelerated ions in their proper reference frame, which is equal to the mass density of undisturbed plasma in the lab frame, $\rho_0=\rho_i\equiv m_i n_i$,  $u^{x}=\gamma_i v_i/c=2\beta_f\gamma_f^2$ is the x-component of the ion four-velocity (the other components $u^{y}= u^{z}=0$), and $\gamma_{i,f}=(1-v_{i,f}^2/c^2)^{-1/2}$ are the gamma-factors. If we assume for simplicity, that the profile of accelerated plasma is a square of side $2a$, then $\Theta(t,\mathbf{r})$ for the laser piston takes  the form
\begin{equation}\label{thetap}
\Theta_p(t,\mathbf{r})=\theta(a-|y|)\theta(a-|z|)\theta(v_i t-x)\theta(x-v_f t)\theta(t)\theta(\tau_p-t),
\end{equation}
where $\theta$ is the Heaviside step function and $\tau_p$ is the piston acceleration time in the lab frame. It is related to the duration of the laser pulse $\tau$ via 
\begin{equation}\label{taup}
\tau_p=\frac{\tau}{1-\beta_f},
\end{equation}
because in the piston frame laser pulse duration is equal to $\tau\sqrt{\frac{1+\beta_f}{1-\beta_f}}$ due to the Doppler effect, and after the transition to the lab frame we obtain (\ref{taup}).

The Fourier transform of the energy-momentum tensor is 
\begin{equation}\label{emftp}
\begin{split}
&\tilde{T}^{xx}(\omega,\mathbf{k})=c\int dtd\mathbf{r}T^{xx}(t,\mathbf{r})e^{i\omega t-i\mathbf{kr}}=\\
&=\frac{16\beta_f^2\gamma_f^4\rho_ic^3\sin k_ya\sin k_z a}{k_xk_yk_z}\\
&\times\left[\frac{e^{i\tau_p(\omega-k_xv_i)}-1}{\omega-k_xv_i}-\frac{e^{i\tau_p(\omega-k_xv_f)}-1}{\omega-k_xv_f}\right].
\end{split}
\end{equation}
In the transverse-traceless gauge the metric distortion caused by a plane gravitational wave can be calculated \cite{Maggiore} as 
\begin{equation}\label{h}
h_{ij}=\frac{4G}{rc^5}\Lambda_{ij,kl}(\mathbf{m})\int\limits_{-\infty}^\infty\frac{d\omega}{2\pi}\tilde{T}^{kl}(\omega,\omega\mathbf{m}/c)e^{-i\omega(t-r/c)},
\end{equation}
where $r$ is the distance to the gravitational wave source, $\mathbf{m}$ is the direction of wave propagation, 
and the Lambda tensor is 
$\Lambda_{ij,kl}(\mathbf{m})=P_{ik}(\mathbf{m})P_{jl}(\mathbf{m})-\frac{1}{2}P_{ij}(\mathbf{m})P_{kl}(\mathbf{m})$ with $P_{ij}(\mathbf{m})=\delta_{ij}-m_i m_j$ 
 the projector with respect to the unit wave vector $\mathbf{m}$ and $\delta_{ij}$ is the Kronecker delta symbol. 

Consider a plane gravitational wave propagating in $z$ direction. The only non--zero components of the Lambda tensor are
$\Lambda_{xx,xx}(\hat{\mathbf{z}})=\Lambda_{yy,yy}(\hat{\mathbf{z}})=\frac{1}{2}, 
\Lambda_{xy,xy}(\hat{\mathbf{z}})=\Lambda_{yx,yx}(\hat{\mathbf{z}})=1,
\Lambda_{yy,xx}(\hat{\mathbf{z}})=\Lambda_{xx,yy}(\hat{\mathbf{z}})=-\frac{1}{2}$.
And then the non-zero components of $h_{ij}$ are $h_{xx}$ and $h_{yy}$ as $\tilde{T}^{yy}=\tilde{T}^{xy}=0$ due to $u^{y}=u^{z}=0$. 

\begin{figure}[h]
\begin{subfigure}{0.45\textwidth}
\includegraphics[scale=0.35]{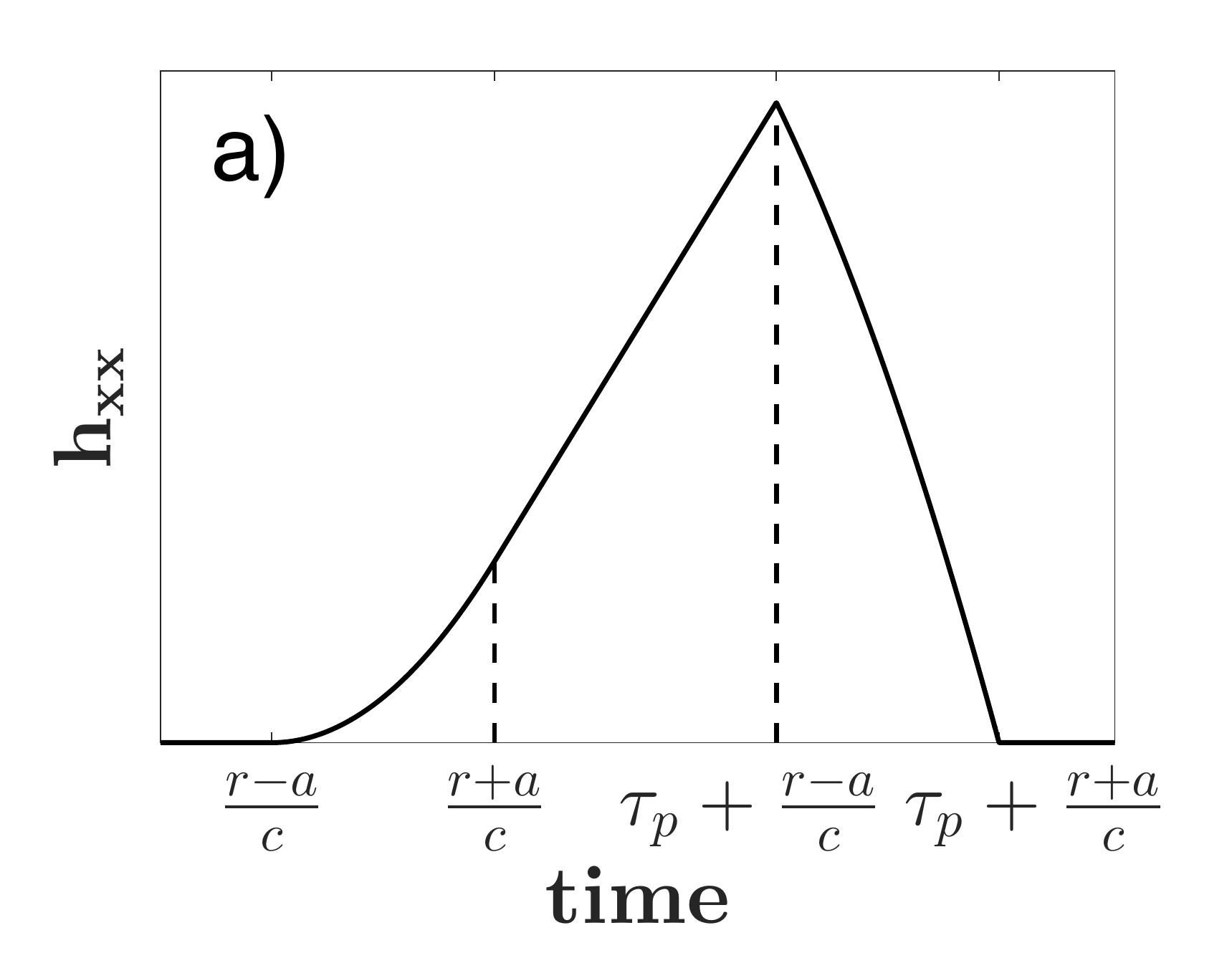}
\end{subfigure}
\begin{subfigure}{0.45\textwidth}
\includegraphics[scale=0.35]{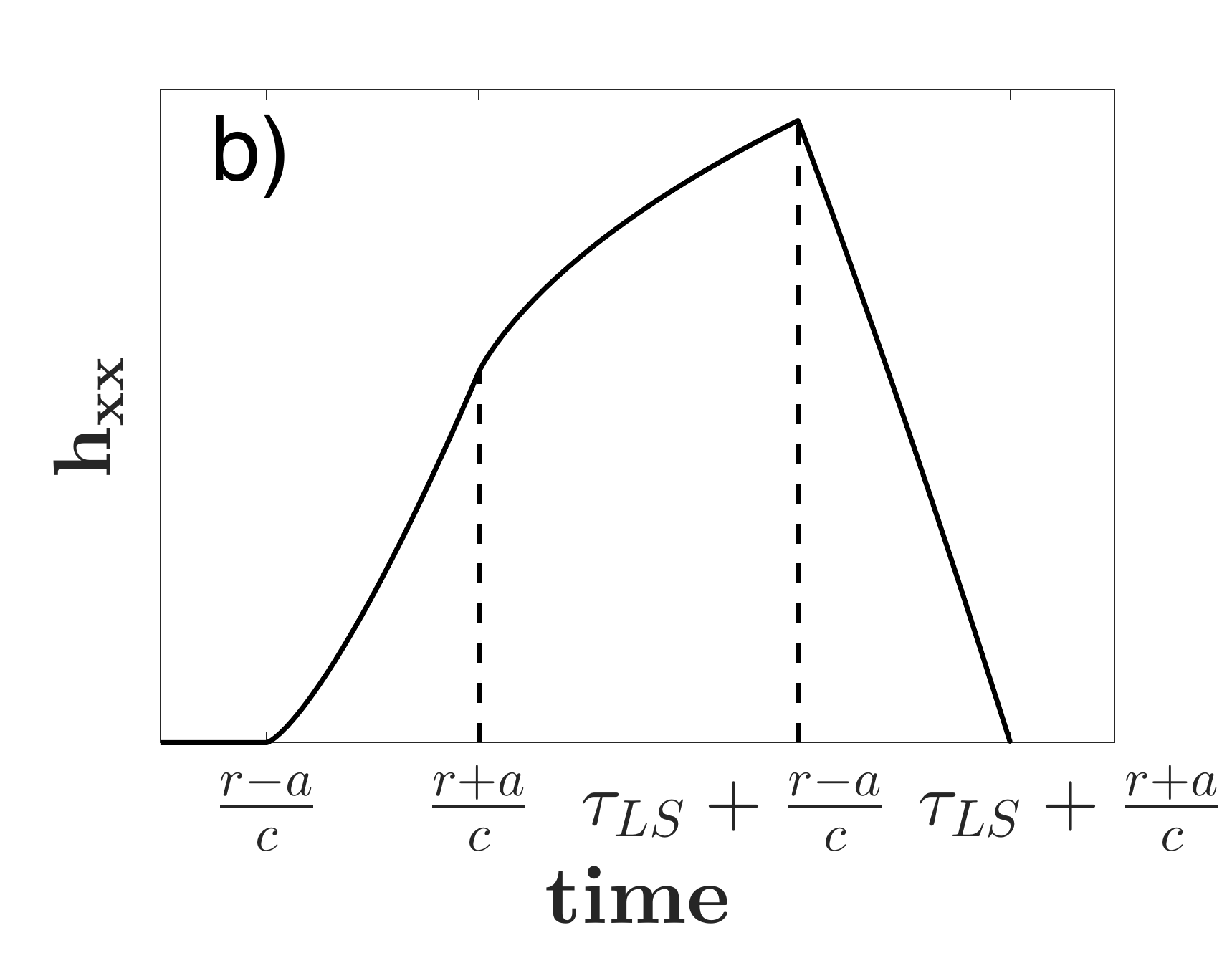}
\end{subfigure}
\caption{The metric perturbation at distance r for: (a) the plasma piston model; (b) the light sail model. In both cases metric perturbation is normalized to $h_{max}$. $\tau_p,\tau_{LS}>2a/c$.}
\label{fig_1}
\end{figure}

In this case, metric distortion takes the form 
\begin{equation}
\label{hp}
h_{xx}=-h_{yy}=\frac{32 G\rho_i \beta_f^2\gamma_f^4(\beta_i-\beta_f) a^3}{\pi r c^2}J(t,r),
\end{equation}
where  
\begin{equation}
\label{J}
J(t,r)=\int\limits_{0}^\infty  d \xi\frac{\sin\xi}{\xi^3}\left[\cos(\mu-\nu)\xi-\cos\nu\xi+\mu\xi\sin(\mu-\nu)\xi\right],
\end{equation}
and $\mu=\frac{c\tau_p}{a}$, $\nu=\frac{ct-r}{a}$. The integral can be calculated explicitly
\begin{equation}
\begin{split}
&J(t,r)=\frac{\pi}{8}\left[|\nu+1|(\nu+1)+|\nu-1|(1-\nu)+\right.\\
&\left.|\mu-\nu+1|(\mu+\nu-1)-|\nu-\mu+1|(\mu+\nu+1)\right],
\end{split}
\end{equation}
and the component of perturbation  $h_{xx}(t)$ is plotted in the Fig.~\ref{fig_1} (a). One can see, that the gravitational wave 
reaches the detector at time $t=(r-a)/c$, when the disturbance from the upper side of the ion target comes to the observation point. The wave leaves the detector at the moment $t=\tau_p+\frac{r+a}{c}$.
$J(t,r)$ reaches its maximum value $c\tau_p/a-1$ at the moment $t=\tau_p+(r-a)/c$, and hence the amplitude of GW in the piston case is equal to
\begin{equation}\label{hmaxp}
h_{max}^{(p)}=\frac{16G\mathcal{E}_p}{rc^4}, 
\end{equation}
where $\mathcal{E}_p=2\rho_i a^2\tau c^3 B^3(1+B)^2/(1+4B+6B^2+4B^3)$ is the total kinetic energy of accelerated ions and we substituted the definitions of $\beta_f$, $\gamma_f$, $\beta_i$ and assumed that $c\tau_p\gg a$. This condition implies that the acceleration time is much longer then the laser period.

According to (\ref{vf}) the shock wave velocity $v_f$ and hence the velocity of accelerated ions $v_i$ increase when $n_i$ decreases. Therefore to maximize the metrics distortion (\ref{hmaxp}) we choose the lowest possible ion density. It is defined by the fact that to make efficient piston acceleration, a nontransparent piston is needed. In the relativistic laser plasma interaction, this transfers into the requirement that the electron density $n_e$ in the piston
is higher that the critical density $n_c$ multiplied by the gamma factor of electrons, which is approximately equal to $a_0$, i.e. for the optimal density we get
\begin{equation}\label{nopt}
\rho_i^{(opt)}= a_0 n_c m_i,
\end{equation}
and hence
\begin{equation}
B^{(opt)}=\sqrt{a_0\frac{m_e}{m_i}}.
\end{equation}

If the distance to the detector is $r=10$ m, laser frequency $\omega=10^{15}$ Hz, laser intensity $I=10^{24}$ W/cm${}^2$, laser pulse duration $\tau=100$ fs, $m_i$ -- proton mass, $a=1$ $\mu$m, then the ions total kinetic energy can be estimated as $\mathcal{E}_p\approx 0.7$ kJ, ions optimal density $\rho_i^{(opt)}\approx 0.6$ g/cm${}^3$ and  maximal metrics distortion $h_{max}^{(p)}\approx9.1\cdot10^{-42}$. 
\subsection{Light sail model}
In the case of the LS-model \cite{Esirkepov2004,Robinson2008} the solution of the equation of
ions motion (\ref{eqls}) reads
\begin{equation}\label{usol}
u^{x}=\sinh\psi-\frac{1}{4\sinh\psi},\quad \psi=\frac{1}{3}\arcsinh(\Omega t+C).
\end{equation}
Here $C=\frac{3(u^{x}_0+\sqrt{(u^{x}_0)^2+1})}{2}+\frac{(u^{x}_0+\sqrt{(u^{x}_0)^2+1})^3}{2}$, $u_0^x$ is the $x$ component of the four velocity of 
ions  at the beginning of the LS acceleration regime  and 
\begin{equation}\label{omega}
\Omega=\frac{6I}{m_{i}n_{i}lc^2}=6\pi\frac{c}{\lambda}\frac{m_e}{m_i}a_0,
\end{equation}
where we took into account estimation (\ref{l}) for target thickness. If we assume, that $\Omega t\gg\max(1,C)$, ($u_{0}^x\ll 1$ and $C \approx 2$) , then the solution (\ref{usol}) can be simplified to
\begin{equation}\label{appsol}
u^{x}\approx\left(\frac{\Omega t}{4}\right)^{1/3},
\end{equation}
because $u^x$ can be approximated as $u^x \simeq \sinh\psi \simeq e^{\psi}/2$ and the  relativistic $\gamma$ factor can be estimated as $\gamma \approx u^x$.

The LS acceleration time can be calculated as (see the explanation after Eq. (\ref{taup}))
\begin{equation}\label{tauls}
\tau_{LS}=\int\limits_0^\tau\frac{dt}{1-v(t)/c}\approx2\int\limits_0^\tau\left(\frac{\Omega t}{4}\right)^{2/3}dt=\frac{6}{5}\left(\frac{\Omega \tau}{4}\right)^{2/3}\tau
\end{equation}
For $I\sim10^{24}$ W/cm${}^2$, $\tau\sim 100$ fs, $m_{i}$ $\sim$ proton mass, $\lambda=1$ $\mu$m, 
$\Omega \tau_{LS}\sim\left(\Omega\tau\right)^{5/3}\sim 10^3\gg1$, and the approximate solution, Eq.~(\ref{appsol}) is valid for almost the whole acceleration time interval $0<t<\tau_{LS}$.

Assume that the profile of the target is a square of side $2a$, and its proper thickness, which is defined by Eq.~(\ref{l}), does not change during the acceleration. The dependence of energy-momentum tensor of accelerated ions (\ref{emt}) on coordinates and time then reads
\begin{equation}\label{thetals}
\begin{split}
\Theta_{LS}(t,\mathbf{r})=&\theta(a-|y|)\theta(a-|z|)\theta(t)\theta(\tau_{LS}-t)\times\\
&\theta(ct+l/\gamma-x)\theta(x-(ct-l/\gamma)).
\end{split}
\end{equation}
After Fourier transform we get
\begin{equation}
\begin{split}
&\tilde{T}^{xx}(\omega,\mathbf{k})=\frac{8\rho_0c^3\sin k_y a\sin k_z a}{k_x k_y k_z}\times\\
&\int\limits_0^{\tau_{LS}} dt \left(\frac{\Omega t}{4}\right)^{2/3} e^{it(\omega-ck_x)}\sin\left[k_x l\left(\frac{\Omega t}{4}\right)^{-1/3}\right].
\end{split}\label{eq:txxSail}
\end{equation} 

Consider GW propagating in $z$ direction. According to the definition Eq.~(\ref{h}), the perturbations in the LS model can be expressed as
\begin{equation}\label{hls}
h_{xx}=-h_{yy}=\frac{6 G\rho_0al}{rc}\left(\frac{\Omega}{4}\right)^{1/3}H(t,r),
\end{equation}
where 
\begin{equation}\label{H}
\begin{split}
&H(t,r)=\theta(ct-r+a)\theta(c\tau_{LS}-ct+r+a)\times\\
&\left[\min\left(t-\frac{r-a}{c},\tau_{LS}\right)^{4/3}-\max\left(0,t-\frac{a+r}{c}\right)^{4/3}\right], 
\end{split}
\end{equation}
and we used the relation $\int\limits_{-\infty}^\infty d\xi\frac{\sin\xi}{\xi}e^{ib\xi}=\pi\theta(1-|b|)$. The function $h_{xx}(t)$ is presented in Fig. \ref{fig_1} (b). If as before $\tau_{LS}\gg2a/c$, the function (\ref{H}) has the maximum
$H_{max}=\frac{8a}{3c}\tau_{LS}^{1/3}$, 
at $t=\tau_{LS}+(r-a)/c$. Hence the amplitude of the gravitational wave takes the form 
\begin{equation}\label{hmaxls}
h^{(LS)}_{max}=\frac{16G\mathcal{E}_{LS}}{rc^4},
\end{equation}
where $\mathcal{E}_{LS}=\left(\Omega\tau_{LS}/4\right)^{1/3}\rho_0a^2lc^2$ is the total energy of accelerated ions. Note, that according to (\ref{l}) $l\sim 1/n_e$, and then GW amplitude in the LS model does not depend on ion density.

%

Considering the values $I=10^{24}$ W/cm${}^2$, $a=\lambda=1\ \mu$m, $\tau=100$ fs, $r=10$ m and $m_{i}$ the mass of proton,  one can estimate $\mathcal{E}_{LS}\approx 0.3$ kJ and the value of (\ref{hmaxls}) as
$h_{max}^{(LS)}\approx 3.7\cdot10^{-42}$. 

The scaling of $h_{max}$ with laser intensity for both models is plotted in the Figure \ref{fig:spec} (a). In all the intensity range considered here, piston model gives higher metric distortion than the light sail model. The same values of $h_{max}$ are reached only at intensities of the order $10^{27}$ W/cm$^2$ but in this range quantum electrodynamic effects must be considered as they may change the interaction significantly. In both models, the scaling with laser intensity is weaker than linear, which means that higher metric distortions can be in principle obtained using larger laser spot size and smaller laser intensity (with the same laser pulse energy). Nevertheless the minimum laser intensity is related to the validity of the piston and the light sail model (radiation pressure acceleration of ions must be dominant) and thus the metric distortion cannot be increased significantly.

\section{Spectrum and angular distribution of GW} 

The spectral angular distribution of GW is given \cite{Maggiore} as:
\begin{equation}\label{spec}
\frac{dE_{GW}}{d\omega dO}=\frac{G\omega^2}{2\pi^2 c^7}\Lambda_{ij,kl}(\mathbf{m})\tilde{T}^{ij}\left(\omega,\frac{\omega\mathbf{m}}{c}\right)\tilde{T}^{kl}
\left(\omega,\frac{\omega\mathbf{m}}{c}\right)^*,
\end{equation}
where $E_{GW}$ is the energy of GW, $O$ is the solid angle and $^{*}$ denotes the complex conjugate. 
In order to eliminate the dependence on a polar angle assume a circular profile of the accelerated plasma, i.e. replace $\theta(a-|y|)\theta(a-|z|)$ in (\ref{thetap}) and (\ref{thetals}) with $\theta(r_\bot-R)$, where $r_\bot=\sqrt{y^2+z^2}$ and $R$ is the radius of ion beam. The Fourier transforms of energy-momentum tensor take the form 
\begin{equation}\label{tcircp}
\begin{split}
&\tilde{T}^{xx}_{p}\left(\omega,\frac{\omega\mathbf{m}}{c}\right)=\frac{8\pi\beta_f^2\gamma_f^4\rho_0Rc^5}{\omega^3\sin\theta\cos\theta}J_1\left(\frac{\omega R\sin\theta}{c}\right)\times\\
&\left[\frac{e^{i\tau_p\omega(1-\beta_i\cos\theta)}-1}{1-\beta_i\cos\theta}-\frac{e^{i\tau_p\omega(1-\beta_f\cos\theta)}-1}{1-\beta_f\cos\theta}\right]
\end{split}
\end{equation}
in the piston case, and
\begin{equation}\label{tcircls}
\begin{split}
&\tilde{T}^{xx}_{LS}\left(\omega,\frac{\omega\mathbf{m}}{c}\right)=\frac{4\pi\rho_0Rc^5\tau_{LS}}{\omega^2\sin\theta\cos\theta}J_1\left(\frac{\omega R\sin\theta}{c}\right)\times\\
&\int\limits_0^1 d\xi\xi^{2/3}e^{i\omega\tau_{LS}(1-\cos\theta)\xi}\sin\left(\frac{\omega l\cos\theta}{c}\left(\frac{\Omega\tau_{LS}\xi}{4}\right)^{-1/3}\right)
\end{split}
\end{equation}
in the Light Sail case. Here $J_1$ is the Bessel function and $\theta$ is the angle between directions of propagation of laser pulse and gravitational wave. 

\begin{figure}[h!]
\includegraphics[scale=0.5]{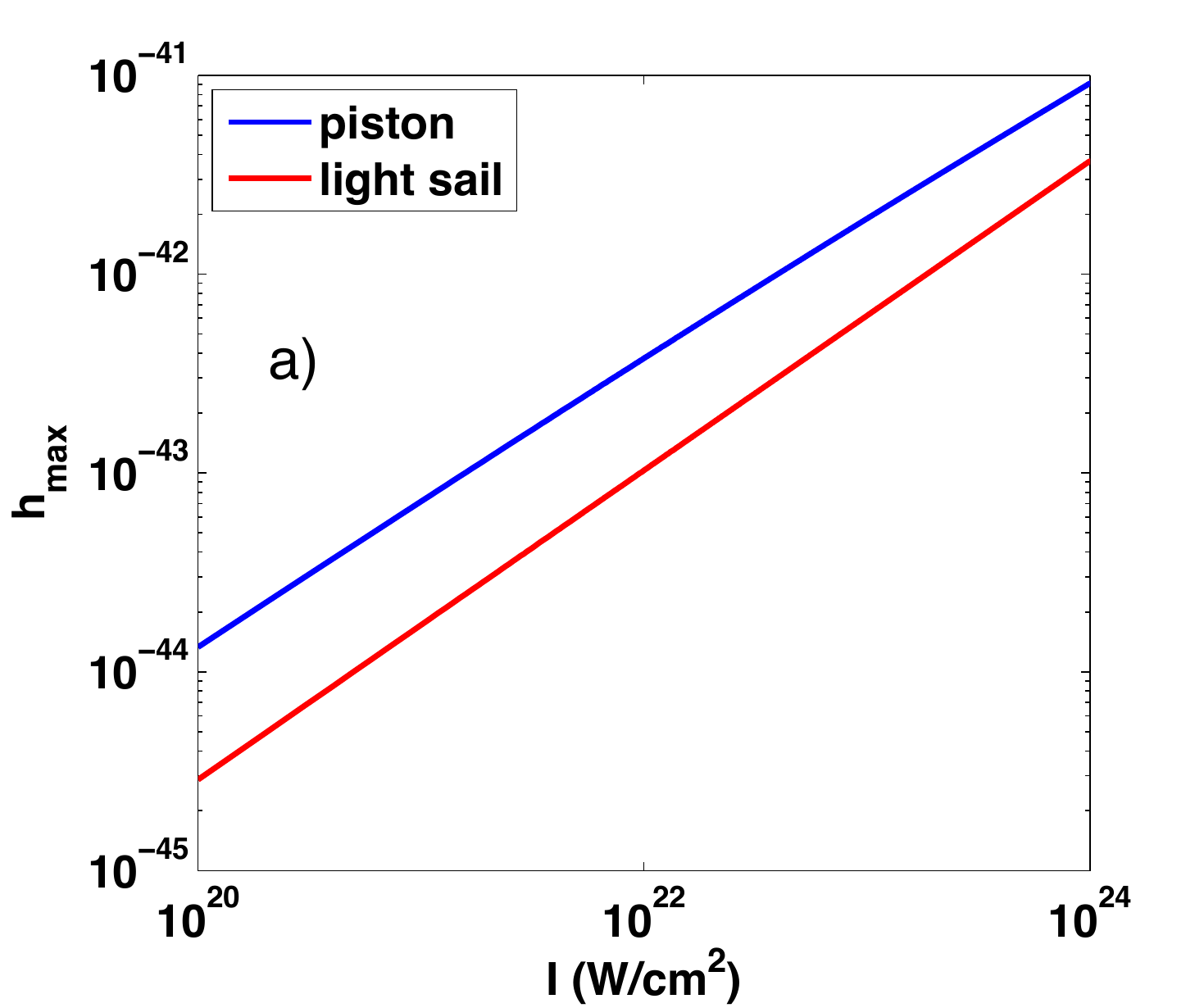}
\includegraphics[scale=0.5]{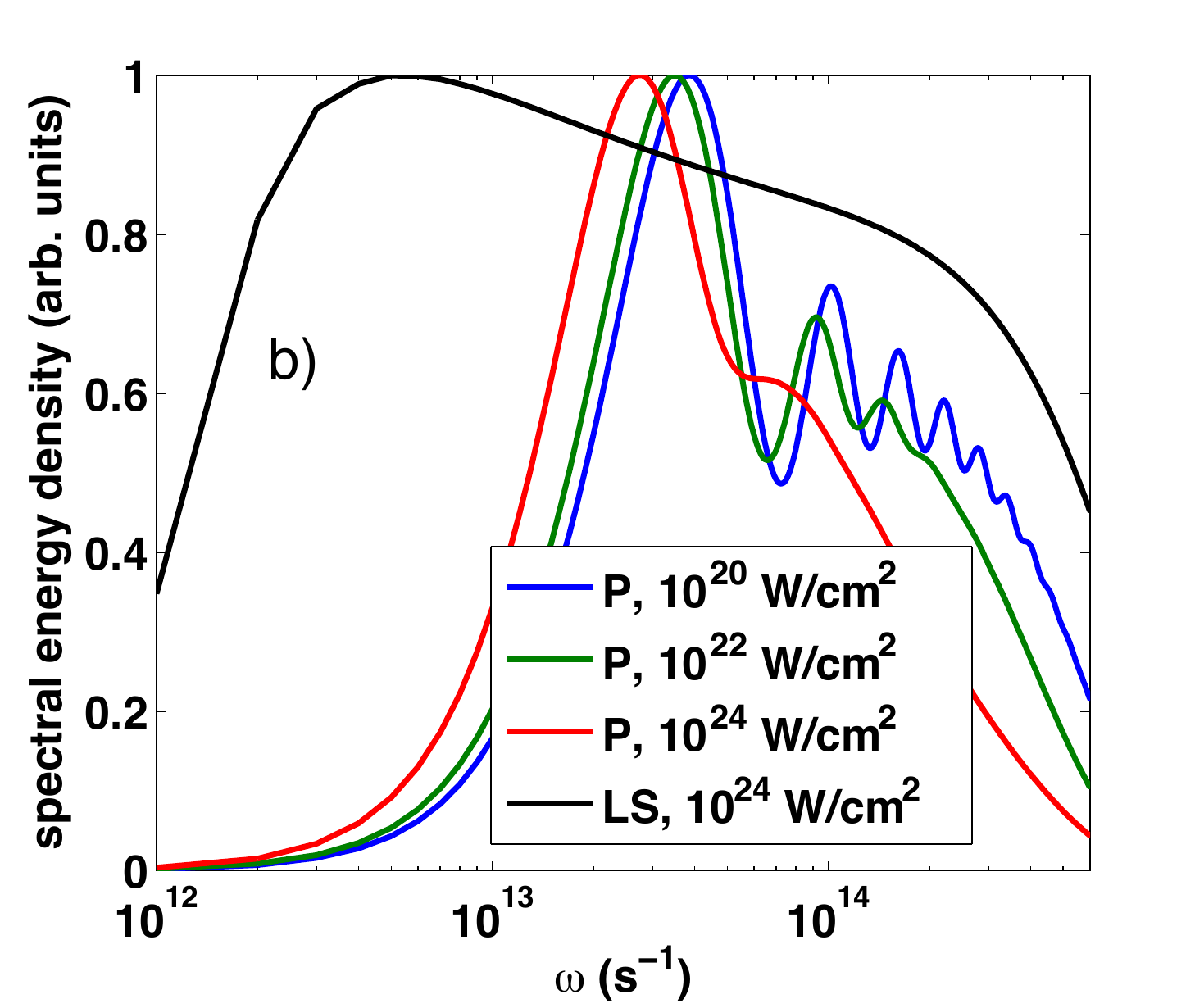}
\caption{(Color online) (a) The scaling of the maximum metric distortions with laser intensity for both models. The parameters are the same like for the spectrum below. (b) The spectrum of gravitational waves in the piston model ('P') for three different intensities.
 The values used are  $R=1\mu$m, $\tau=100$ fs and the density $\rho_0$ is given by (\ref{nopt}). The spectrum for light sail model ('LS') and the same parameters is also included, but $\rho_0=1$ g/cm$^3$ and $l$ is given by (\ref{l}) in this case. In both cases the spectra are normalized to $(dE/d\omega)_{max}$.
}\label{fig:spec}
\end{figure}

Taking into account that $\Lambda_{xx,xx}=\sin^4\theta/2$ and integrating (\ref{spec}) over solid angle we get the spectrum of GW, which is presented on the Figure \ref{fig:spec} (b). The maximum of the spectrum is located in the region $\omega\sim 1/\tau_{p,LS}$ because the perturbation is not periodic and has a finite duration. We observe, that the spectrum shifts to the lower frequencies with the growth of laser pulse intensity because of relativistic increasing of acceleration time, see Eqs.~(\ref{taup}), (\ref{tauls}).

Note, that for pulse parameters under consideration and target size of the order of microns $\omega a/c\sim\omega R/c\ll 1$ if the frequency $\omega$ is located in the part of spectrum, where GW emission is significant, see figure \ref{fig:spec} (b).  Therefore sine and Bessel functions in the equations (\ref{emftp}), (\ref{eq:txxSail}), (\ref{tcircp}) and (\ref{tcircls}) can be expanded into Taylor series up to the first term, and Fourier transforms for energy-momentum tensor for circular and square targets become the same. It means, that the spectrum and also the angular distribution of gravitational radiation do not depend on the shape of the target.

\begin{figure*}
\includegraphics[scale=0.4]{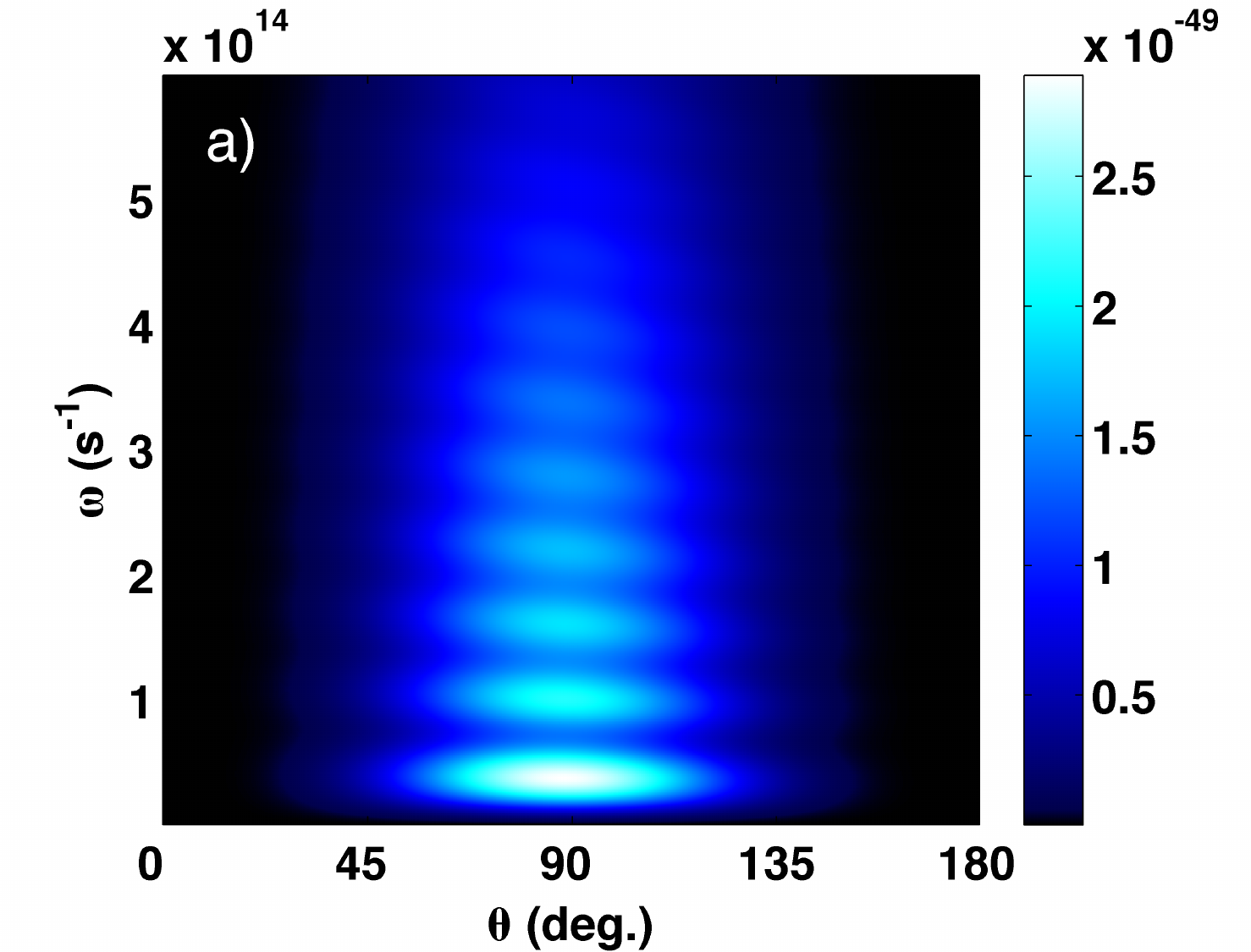}
\includegraphics[scale=0.4]{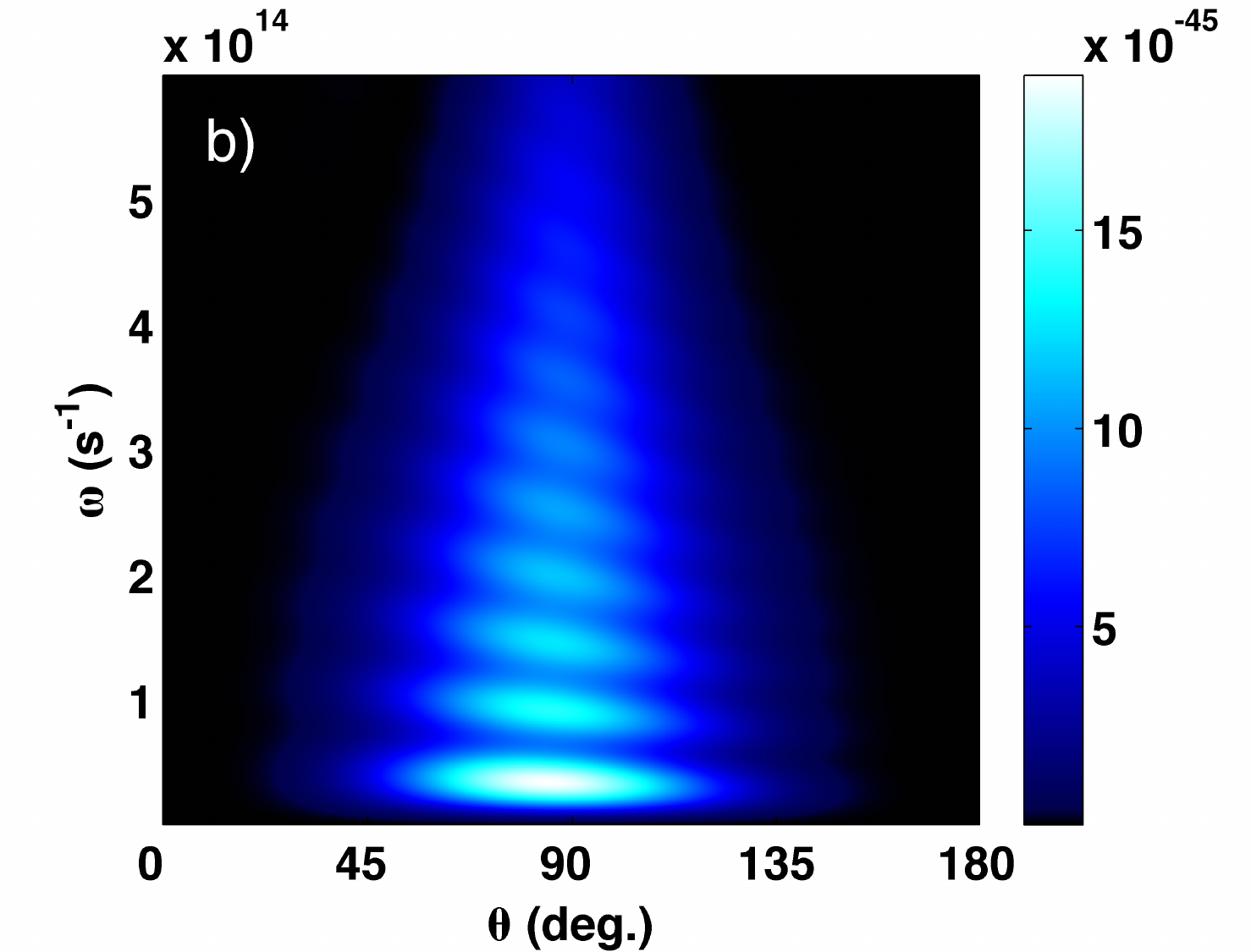}
\includegraphics[scale=0.4]{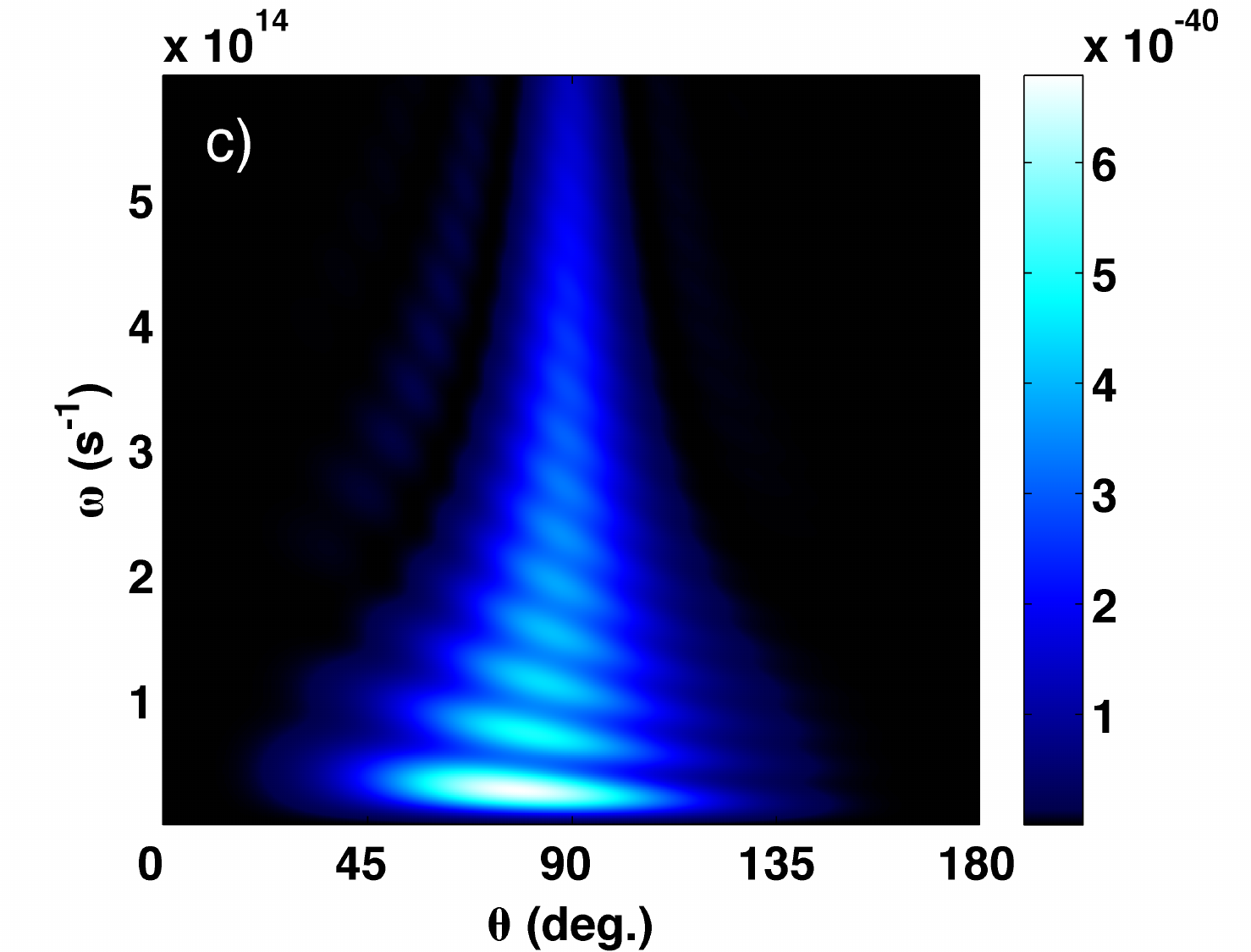}
\includegraphics[scale=0.4]{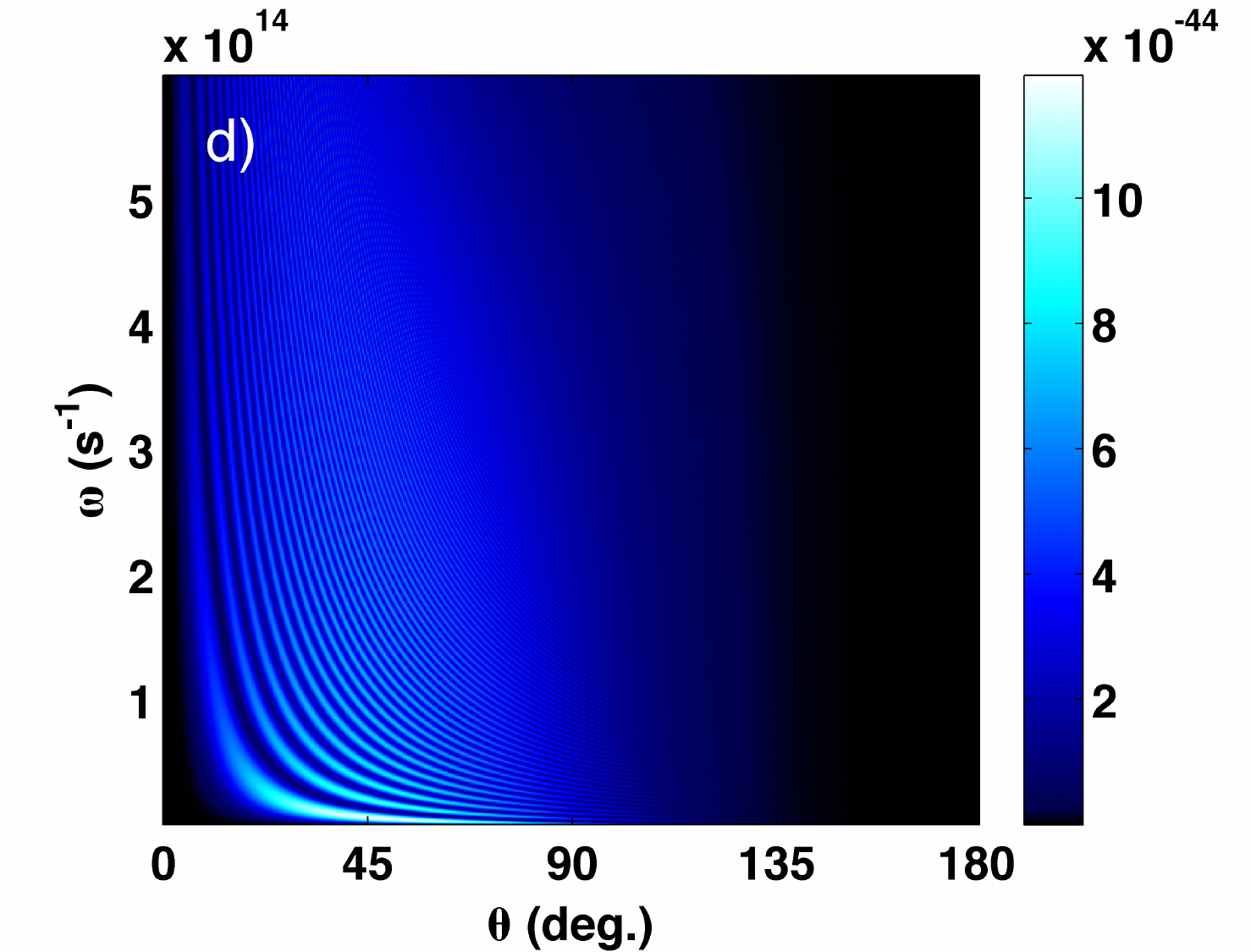}
\caption{(Color online) (a) -- (c) The angular distribution $dE_{GW}/(d\omega d\theta)$ of gravitational radiation in the piston model for three different intensities: (a) the non-relativistic piston at $I=10^{20}$ W/cm${}^2$, (b) the weakly 
relativistic piston at $I=10^{22}$ W/cm${}^2$, (c) the relativistic one for $I=10^{24}$ W/cm${}^2$. (d) The angular distribution $dE_{GW}/(d\omega d\theta)$ of gravitational radiation in the light sail model for  $I=10^{24}$ W/cm${}^2$.
The angle $\theta$ is measured with respect to the direction of laser propagation,  $\theta=0^{\circ}$ on $x$ axes. Other parameters are the same like in figure \ref{fig:spec}.
}\label{fig:angle}
\end{figure*}

The angular distribution $dE_{GW}/(d\omega d\theta)$ of the gravitational radiation is visualised in the Figure \ref{fig:angle}. We observe the alignment
of the radiation with the direction of propagation as ions become relativistic. Indeed, (a) panel corresponds to $v_i\sim 0.11c$, (b) corresponds to $v_i\sim 0.3c$, (c) corresponds to $v_i\sim 0.65c$ and (d) corresponds to final ions velocity $v_i\sim 0.994c$ .

\section{Conclusion}
The generation of high-frequency gravitational waves by high-power laser systems was considered and their functional dependence on 
the laser-plasma interaction parameters was derived.

In the piston regime the source of GW is the shock wave in a thick plasma
target, continuously increasing the mass of accelerated matter. In the light sail regime the mass of accelerated matter remains constant, but the velocity is increasing. 

The spectrum of GW has a typical pulse-like form with the maximum at $(dE/d\omega)_{max}\sim 1/\tau_{GW}$, where $\tau_{GW}$ is the duration of GW. For realistic laser and plasma parameters the GW frequency is of the order of tens of THz. The direction of GW propagation depends on the velocity of ions. Non-relativistic ions emit GW transversely to the direction of acceleration, and the direction of GW propagation turns to the direction of acceleration with the increasing of ion velocity.

In both cases investigated, the perturbations of the space-time metric are small and detection is a challenge. 
High-frequency GW detectors were suggested \cite{baker2008,li2009,fang2013}, which are based on the coupling between gravitational and 
electromagnetic wave in the presence of background magnetic field (inverse Gertsenshtein effect \cite{gertsenshtein1962}). 
The sensitivity for such detector was estimated to be of the order of $h_{rms}\sim10^{-34}$~Hz${}^{-1/2}$, see \cite{baker2008}. For GW under consideration in the present paper $h_{rms}\sim h\sqrt{\tau_{GW}}\sim10^{-47}$~Hz${}^{-1/2}$.

It should be noted, that the GW amplitude could not be sufficiently increased by increasing of the laser intensity, because instead of acceleration the laser energy would be wasted on strong field QED processes such as QED cascades \cite{Bell2008}, which start at $I\gtrsim 10^{25}$ W/cm${}^2$ or electron-positron pair creation from vacuum \cite{Schwinger1951,Bulanov2010}, which starts at $I\gtrsim 10^{27}$ W/cm${}^2$ and in principle is able to totally exhaust the laser pulse \cite{Fedotov2010}. The focus spot radius of high intensity lasers is of the order of wavelength, and it additionally restricts GW amplitude, due to limiting the number of accelerated ions. However increasing of the laser pulse energy (i.e. the size of the focus spot and the pulse duration) is a way to increase the energy of accelerated ions and hence GW amplitude. 

E. G., H. K., S. W and G. K were supported by ELI project No.CZ.02.1.01./0.0/0.0/15-008/0000162.
O.K. acknowledges support from the Czech Science Foundation Project 15-02964S.

\end{document}